\documentclass[%
reprint,
superscriptaddress,
showpacs,
nofootinbib,
amsmath,amssymb,
prc,
floatfix, ]%
{revtex4-1}

\usepackage{color}

\usepackage{graphicx}
\usepackage{dcolumn}
\usepackage{bm}
\usepackage[dvipdfmx,bookmarks=true,colorlinks,%
            citecolor=blue,linkcolor=blue,anchorcolor=blue,filecolor=blue,urlcolor=blue,%
           ]{hyperref}

\allowdisplaybreaks

\newcommand{\bra}[1]{\langle {#1} |}
\newcommand{\ket}[1]{| {#1} \rangle}

\begin{document}


\title{
Collective inertial masses in nuclear reactions}

\author{Kai Wen}%
 \email{wenkai@nucl.ph.tsukuba.ac.jp}
 \affiliation{Center for Computational Sciences,
              University of Tsukuba, Tsukuba 305-8577, Japan}

\author{Takashi Nakatsukasa}%
 \affiliation{Center for Computational Sciences,
              University of Tsukuba, Tsukuba 305-8577, Japan}
 \affiliation{Faculty of Pure and Applied Sciences,
              University of Tsukuba, Tsukuba 305-8571, Japan}
 \affiliation{iTHES Research Group, RIKEN, Wako 351-0198, Japan}

\date{\today}

\begin{abstract}

Towards the microscopic theoretical description for large amplitude
collective dynamics, we calculate the coefficients of
inertial masses for low-energy nuclear reactions.
Under the scheme of energy density functional,
we apply the adiabatic self-consistent
collective coordinate (ASCC) method,
as well as the Inglis' cranking formula to calculate the
inertias for the translational and the relative motions,
in addition to those for the rotational motion.
Taking the scattering between two $\alpha$ particles as an example,
we investigate the impact of the time-odd
components of the mean-field potential on the collective inertial masses.
The ASCC method asymptotically reproduces the exact masses for both
the relative and translational motions.
On the other hand,
the cranking formula fails to do so when the time-odd components exist.
\end{abstract}


\maketitle

\section{Introduction}
The time-dependent density functional theory (TDDFT)
\citep{Neg82,Sim12,Nak12,MRSU14,NMMY16}
is a general microscopic theoretical framework to study low-energy nuclear reactions.
Based on the TDDFT, the mechanisms
of nuclear collective dynamics have been extensively studied for decades.
The linear approximation of TDDFT leads to the random-phase
approximation (RPA) \citep{RS80,BR86,NMMY16}, which is capable of
calculating nuclear response functions and providing us a unified description
for both structural and dynamical properties.
Despite the detailed microscopic information revealed by TDDFT,
it has a difficulty in describing nuclear collective dynamics at low energy
\citep{NMMY16}.
For instance, it cannot describe the
sub-barrier fusion and spontaneous fission, due to its
semiclassical nature \citep{RS80,Neg82,NMMY16}.

The description of nuclear dynamics in terms of collective degrees of freedom
has been explored in nuclear reaction theories.
However, the derivation of the ``macroscopic'' reaction model
based on the microscopic nuclear dynamics has been rarely studied
in the past.
For the theoretical description in terms of
collective degrees of freedom, the collective inertial masses
with respect to the collective coordinates
are of paramount importance.
One of the most commonly used methods to extract the collective mass
coefficient is the Inglis' cranking formula \citep{BK68-1,BK68-2,YLQG99},
which can be derived based on the
adiabatic perturbation theory.

It is well-known that the cranking formula has a problem that
it fails to reproduce the total mass for the translational motion
of the center of mass of a nucleus \cite{RS80}.
Therefore, it is highly desirable to replace the cranking mass
by the one theoretically more advanced and justifiable.
We believe that the
adiabatic self-consistent collective coordinate (ASCC)
method \citep{MMSK80,MNM00,HNMM07,HNMM09} suites for this purpose.
The method, in the first place, aims at determining the
canonical variables on the optimal collective subspace
for description of a low-energy collective motion.
The masses with respect to those collective coordinates can be extracted
by solving a set of the ASCC equations.
This method has been applied to many nuclear
structure problems with large-amplitude nuclear dynamics
with the Hamiltonian of
the separable interactions \citep{HNMM07,HNMM09,HLNNV12,SHYNMM12}.
Recently, by combining the imaginary-time
evolution \citep{DFKW80} and
the finite amplitude method \citep{NIY07,INY09,AN11,AN13},
we proposed a numerical method to solve the ASCC equations and
to determine the optimal collective path for nuclear
reaction \citep{WN16}.
At the same time, we obtain the collective inertial mass
in a self-consistent manner.
In this work, we calculate the collective masses for three modes of
collective motion,
the translational motion,
the relative motion and rotational motion.
We compare the ASCC results
with those of the cranking formula.

Our calculations are under the scheme of energy density functional theory.
In order to guarantee the Galilean symmetry during a collective motion,
most of the energy density functionals must include
densities that are odd with respect to the time reversal.
Under the assumptions of the time-reversal symmetry,
these terms vanish and therefore do not
contribute to the time-even states,
while they have non-zero values in situations of dynamical reactions.
It has been found that the time-odd components play an important role
in the inertia parameters for nuclear rotations\citep{EBR16,Li12}.
To investigate this problem in the context of reaction dynamics
of light nuclei,
we investigate the effects of time-odd terms on the
different inertial masses,
taking the $\alpha+\alpha$ reaction as the simplest example.

This paper is organized as the following. In Sec.~\ref{sec:theo}, we
recapitulate the formulation of the basic ASCC
equations in the case of one-dimensional collective motion.
We present the method of constructing the collective path and
the coordinate transformation procedure to calculate the inertial mass
parameter with respect to the relative coordinate.
In Sec.~\ref{sec:app}, 
we apply the method to the
reaction system $\alpha$+$\alpha$$\leftrightarrow$$^{8}$Be.
We focus on the influence of the time-odd terms
on both the relative and rotational inertias.
Summary and concluding remarks are
give in Sec. \ref{sec:sum}.

\section{Theoretical framework}\label{sec:theo}
\subsection{Formulation of ASCC method}
In this section, neglecting the paring correlation,
we recapitulate the basic ASCC formulation,
and introduce the numerical procedure of constructing the collective
path and calculating the inertial mass.
The details can be found in Ref.~\citep{WN16}.

For simplicity, here we consider the collective motion described by
only one collective coordinate $q(t)$, which has a
conjugate momentum $p(t)$.
We assume that the time-dependent mean-field states
are parameterized by
Slater determinants labeled as $|\psi(p,q)\rangle$.
The energy of the system reads
\begin{eqnarray}
H(p, q)= \langle \psi(p, q)|\hat{H} |\psi(p, q)\rangle, \label{pdef}
\end{eqnarray}
which defines a classical collective Hamiltonian.
In the ASCC method,
the resulting collective path $|\psi(p,q)\rangle$ is determined so as to
maximally be decoupled from other intrinsic degrees of freedom.
The evolution of $q(t)$ and $p(t)$
obeys the canonical equations of motion
with the classical Hamiltonian $H(p, q)$.

In order to consider the adiabatic limit,
we assume the momentum $p$ is small and
the states are expanded in powers of $p$ about $p = 0$.
The states $|\psi(p,q)\rangle$ are written as
\begin{eqnarray}
|\psi(p,q)\rangle = e^{i p \hat{Q}(q)}|\psi(0,q)\rangle
	= e^{i p \hat{Q}(q)}|\psi(q)\rangle,
\label{eq1-2}
\end{eqnarray}
where the generator $\hat{Q}(q)$ is defined as
$\hat{Q}(q)|\psi(q)\rangle = -i \partial_{p}|\psi(q)\rangle$.
The conjugate $\hat{P}(q)$ is introduced as a
generator for the infinitesimal translation in $q$,
$\hat{P}(q)|\psi(q)\rangle = i \partial_{q}|\psi(q)\rangle$.
$\hat{P}(q)$ and $\hat{Q}(q)$
can be expressed in the form of one-body operator as
\begin{eqnarray}
\hat{P}(q)&=&i \sum_{n\in p,j\in h}P_{nj}(q)a^{\dagger}_{n}(q)a_{j}(q)
+ \mathrm{h.c.}, \quad \nonumber\\
\hat{Q}(q)&=&\sum_{n\in p,j\in h}Q_{nj}(q)a^{\dagger}_{n}(q)a_{j}(q)
+ \mathrm{h.c.},
 \label{Q(q)}
\end{eqnarray}
where $i$ in the expression of $\hat{P}(q)$ is simply for convenience.
They are locally defined at each coordinate $q$ and will change their
structure along the collective path.
The particle ($n\in p$) and hole ($j\in h$) states are also defined with
respect to the Slater determinant $|\psi(q)\rangle$.

In the adiabatic limit,
expanding Eq. (\ref{eq1-2}) 
with respect to $p$ up to second order,
the invariance principle of the self-consistent collective coordinate
(SCC) method \citep{MMSK80}
leads to the equations of the
ASCC method \citep{MNM00,NMMY16}.
Neglecting the curvature terms, it reduces to somewhat simpler equation set:
\begin{eqnarray}
&&\delta\langle \Psi(q)
|\hat{H}_{\rm mv}|
\Psi(q)\rangle = 0, \label{chf}\\
&&\delta\langle \Psi(q)|[\hat{H}_{\rm mv},\frac{1}{i}\hat{P}(q) ]
        - \frac{\partial^{2} V(q)}{\partial q^{2}} \hat{Q}(q)
        |\Psi(q)\rangle = 0, \label{RPA0} \\
&&\delta\langle \Psi(q)|[\hat{H}_{\rm mv},i\hat{Q}(q)]
         - \frac{1}{M(q)}\hat{P}(q)   |\Psi(q)\rangle = 0, \label{RPA}
\end{eqnarray}
with the inertial mass parameter $M(q)$.
The mass $M(q)$ depends on the scale of the coordinate $q$.
Thus, we can choose it to make $M(q)=1$ without losing anything.
The moving mean-field Hamiltonian $\hat{H}_{\rm mv}$
and the potential $V(q)$ are respectively defined as
\begin{eqnarray}
\hat{H}_{\rm mv}=\hat{H}- \frac{\partial V(q)}{\partial q} \hat{Q}(q),\quad
V(q)= \langle \psi(q)|\hat{H} |\psi(q)\rangle. \label{pdef}
\end{eqnarray}
Note that the collective path is given by $\ket{\psi(q)}$, which
represents the state $\ket{\psi(q,p)}$ with $p = 0$.
Equation (\ref{chf}) is similar to a constrained
Hartee-Fock problem, however, the
constraint operator $\hat{Q}(q)$ depends on
the coordinate $q$, which is self-consistently
determined by the RPA-like equations (\ref{RPA0}) and (\ref{RPA}),
called ``moving RPA equations''.
The conventional RPA forward and backward amplitude $X_{ni}(q)$ and
$Y_{ni}(q)$ can be regarded as the linear combination of $\hat{P}(q)$ and
$\hat{Q}(q)$.
\begin{eqnarray}
   X_{nj} &=&\sqrt{\frac{\omega}{2}}Q_{nj}
     +\frac{1}{\sqrt{2\omega}}P_{nj}, \quad \nonumber\\
   Y_{nj} &=&\sqrt{\frac{\omega}{2}}Q_{nj}
     -\frac{1}{\sqrt{2\omega}}P_{nj},
\end{eqnarray}
where the RPA eigenfrequency $\omega$ is related to the mass parameter
and the second derivative of the potential
\begin{eqnarray}
\omega^{2}=\frac{1}{M(q)}\frac{\partial^{2} V(q)}{\partial q^{2}}.
\label{ab4}
\end{eqnarray}

As a pair of canonical variables, a weak canonicity condition
$\langle\Psi(q)|[i\hat{P}(q),\hat{Q}(q)]|\Psi(q)\rangle=1 \label{weak}$
should be satisfied.
This canonicity condition is automatically satisfied if
the RPA normalization condition $\sum_{n,j}(X_{nj}^{2}-Y_{nj}^{2})=1$
holds.

It should be noted that the ASCC method is applicable to
systems with pairing correlations, in principle.
However, in this paper, we neglect the pairing correlation to reduce
the computational cost, and concentrate our discussion on effects of
mean fields of particle-hole channels for the inertial masses.
We present results for the $\alpha + \alpha$ reaction in
Sec. \ref{sec:app}, for which no level crossing at the Fermi
surface is involved. Therefore, the pairing plays very little
role in this particular case.

For superconducting systems, apart from the collective coordinate and
momentum, an additional pair of canonical variables, the particle
number and the conjugate gauge angle, are needed to label the nuclear state.
Details of the formulation are give in Refs.~\citep{MNM00,NMMY16}.

\subsection{ASCC collective path and inertial mass}\label{sec:theo2}
A change in the scale of the collective coordinate $q$ results
in a change in the collective mass $M(q)$.
Thus, in order to discuss the magnitude of the collective mass,
we need to fix its scale.
This is normally done by adopting an intuitive choice of the
one-body time-even operator $\hat{O}$.
One of possible choices is the mass quadrupole operator
$Q_{20}=\int d\mathbf{r}
\psi^\dagger(\mathbf{r})r^2Y_{20}(\hat{\mathbf{r}})\psi(\mathbf{r})$.
In the present study of nuclear scattering (nuclear fission),
it is convenient to adopt the relative distance $\hat{R}$ between two nuclei
with the projectile mass number $A_\mathrm{pro}$
and the target mass number $A_\mathrm{tar}$.
Assuming that the center of mass of the two nuclei are
on the $x$ axis ($y=z=0$),
\begin{eqnarray}
\hat{R}\equiv \int d\mathbf{r} \hat{\psi}^\dagger(\mathbf{r})\hat{\psi}(\mathbf{r})
             x\left[\frac{\theta(x-x_{\rm s})}{A_{\rm pro}}-
         \frac{\theta(x_{\rm s}-x)}{A_{\rm tar}}\right]  ,\label{defr}
\end{eqnarray}
where $\theta(x)$ is the step function,
and $x=x_{\rm s}$ is the artificially introduced
section plane that divides the total space
into two, each of which contains the nucleon number of
$A_{\rm pro}$ and $A_{\rm tar}$, respectively.

The operator $\hat{R}$ has an evident physical meaning
when the projectile and the target are far away to each other.
When they touch each other, the distance between two nuclei
is no longer a well-defined quantity, thus loses its significance.
However, this is not a problem in the present microscopic formulation of the
reaction model.
We have determined the reaction path and the canonical variables $(q,p)$,
through the ASCC method.
It is merely a coordinate transformation from $q$ to $R$
with a function $R(q)$.
The reaction dynamics do not depend on the choice of $R$,
as far as the one-to-one correspondence between $q$ and $R$ is valid.

The coordinate transformation naturally leads to
the transformation of the inertial mass from $M(q)$ to $M(R)$;
\begin{eqnarray}
M(R) =M(q)\left(\frac{dq}{dR}\right)^{2}. 
\label{mass}
\end{eqnarray}
The calculation of the derivative $dq/dR$ is straitforward,
because the collective path $\ket{\psi(q)}$ and
the local generator $\hat{P}(q)$ of the coordinate $q$ are obtained
by solving the ASCC equations (\ref{RPA0}) and (\ref{RPA}).
\begin{eqnarray}
\left(\frac{dq}{dR}\right)^{-1} &=& \frac{dR}{dq}
	= \frac{d}{dq}\bra{\psi(q)}\hat{R}\ket{\psi(q)} \nonumber\\
&=& -i \bra{\psi(q)}\left[\hat{R},\hat{P}(q)\right] \ket{\psi(q)} .
\end{eqnarray}
The inertia mass parameter with respect to $R$ or any other coordinate
can be easily calculated with this formula.

We solve the moving RPA equations (\ref{RPA0}) and (\ref{RPA})
by taking advantage of
the finite amplitude method (FAM) \citep{NIY07,INY09,AN11,AN13},
especially the matrix FAM prescription \citep{AN13}.
To solve the ASCC equations (\ref{chf}), (\ref{RPA0}), and (\ref{RPA})
self-consistently and
construct the collective path $\ket{\psi(q)}$,
we adopt the following procedures:
\begin{enumerate}
\item
	Prepare the Hartree-Fock ground state $\ket{\psi(q=0)}$
	which can be either the two separated nuclei before fusion,
	or the ground state of the mother nucleus before fission.
\item
	Based on $\ket{\psi(q)}$, solve the moving RPA equations
	(\ref{RPA0}) and (\ref{RPA}),
	to obtain $\hat{Q}(q)$ and $\hat{P}(q)$.
	First, we start with an axpproximation $\hat{Q}(q+\delta q)=\hat{Q}(q)$.
\item
	\label{step3}
	Solve the moving HF equation (\ref{chf}) to calculate
	the state $\ket{\psi(q+\delta q)}$
	by imposing the condition
\begin{eqnarray}
\langle\Psi(q+\delta q) |\hat{Q}(q)| \Psi(q+\delta q)\rangle
 = \delta q,
\label{cons}
\end{eqnarray}
where we use the approximate relation,
$\ket{\psi(q+\delta q)}\simeq e^{-i\delta q \hat{P}(q)}\ket{\psi(q)}$,
to constrain the step size.
\item
	\label{step4}
	With this new state $\ket{\psi(q+\delta q)}$,
	update the generators
	$\hat{Q}(q+\delta q)$ and $\hat{P}(q+\delta q)$ by solving the
	 moving RPA equations again.
	Then, with these updated generators, go back to the step \ref{step3}.
	Repeat the steps \ref{step3} and \ref{step4}
	until the self-consistency is achieved at $q+\delta q$.
\item
	Then, regarding $q+\delta q$ as $q$ with an initial approximation
	$\hat{Q}(q+\delta q)=\hat{Q}(q)$,
	go to the step \ref{step3}.
\end{enumerate}
Carrying on this iterative procedure, we determine a series of states
$\ket{\psi(0)}, \ket{\psi(\delta q)}, \ket{\psi(2\delta q)},
\ket{\psi(3\delta q)},\cdots$ that form
the ASCC collective path.
Changing the sign of the right hand side of Eq. (\ref{cons}),
we can also construct the collective path toward the opposite direction
$\{ \ket{\psi(-\delta q)}, \ket{\psi(-2\delta q)}, \cdots\}$.
In this way, the collective path $\ket{\psi(q)}$, the potential $V(q)$,
and the collective mass $M(q)$ are determined
self-consistently and no \textit{a priori} assumption is used.

\section{Applications }
\label{sec:app}
\subsection{Solutions for the translational motion }
\label{sec:alpha}

First, we calculate the inertial mass for the translational motion,
for which we know the exact value $Am$.
The calculation is done in the three-dimensional
coordinate space discretized in the square grid
in a sphere with radius equal to 7 fm. 
The BKN energy density functional \citep{BKN76} is adopted in the present calculation.

\begin{figure}
\begin{centering}
\includegraphics[height=5.5cm]{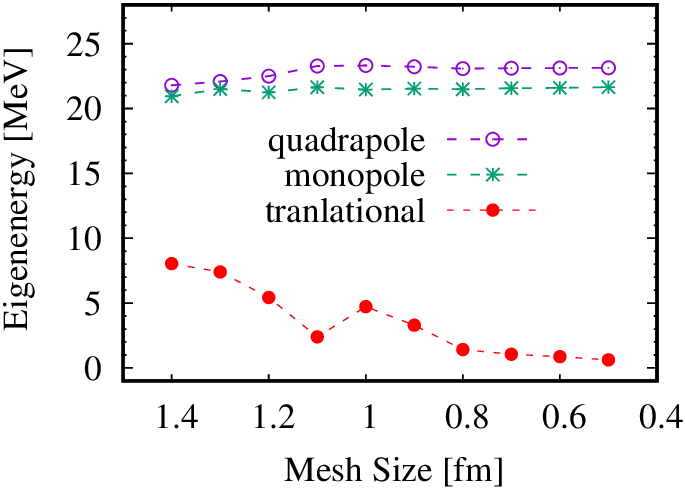}\quad\includegraphics[height=5.99cm]{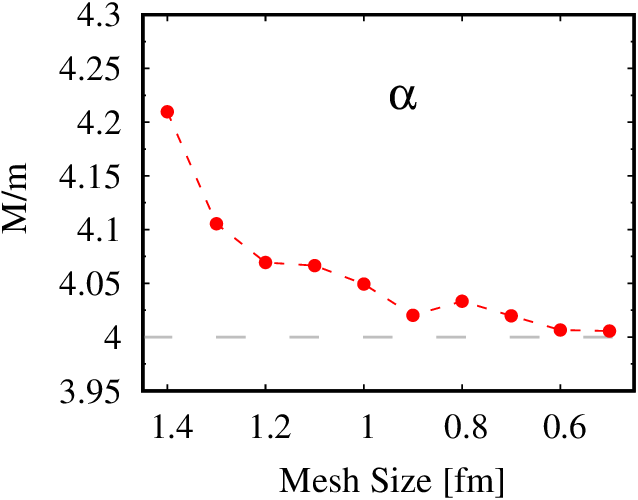}
\caption{Left panel:
Calculated RPA eigenfrequencies based on the ground state of the alpha particle,
as a function of mesh size.
The red closed circles indicate the values for translational mode, while
the green asterisks and purple open circles
indicate those for monopole and quadrupole modes, respectively.
Right panel:
Calculated translational mass $M$ of a single alpha particle in units of
nucleon's mass $m$, as a function of the mesh size.
The calculated mass with respect to the $y$ direction, perpendicular
to the symmetry ($x$) axis, is shown. The one with respect to the $x$
direction is presented in Fig. 1 of reference \citep{WN16}.
}
\label{fig:translation}
\end{centering}
\end{figure}

The HF ground state is a trivial solution of Eqs. (\ref{chf}), (\ref{RPA0}), and (\ref{RPA}),
on the collective path since it corresponds to
the minimum of the potential surface, $\partial V/\partial q=0$.
We calculate the translational
inertia mass of the ground state of an alpha particle, and
examine its grid size dependence.
The left panel of Fig. \ref{fig:translation} shows the eigenfrequency
$\omega$ in Eq. (\ref{ab4}) of the lowest several RPA states
as a function of the mesh size of the grids.
The three translational
modes along $x, y, z$ axis are degenerated and shown by the red
dots, the absolute value of this eigenfrequency decreases and
approaches zero as the mesh size becomes smaller.
The value of the translational motion is significantly smaller than
all the other collective modes.
In the ideal case
where the mesh size is sufficiently small, this value is expected to be zero.
For other collective modes, the eigenfrequencies
stay almost constant as functions of the mesh size.
Due to the compact nature of alpha particle, except for
the translational zero-modes, the lowest physical excitation mode is
calculated to be about
$20$ MeV, which represents the monopole vibration.

Using Eq. (\ref{mass}) we
calculate the translational inertia mass of one alpha particle.
The right panel of Fig. \ref{fig:translation} shows the result
as a function of mesh size.
As the mesh size decreases, the results approach to the value of 4
in the unit of nucleon mass, which is the exact total mass of
the alpha particle.
With the simple BKN energy density functional, this exact value for the translation is
also obtained with the cranking mass formula of Inglis.
However, it underestimates the exact total mass when the energy functional
has a effective mass $m^*/m<1$.
On the other hand, the ASCC mass for the translational motion
is invariant and exact even with the effective mass.
This is due to the Galilean symmetry of the energy density functional
which inevitably contains the time-odd components.
This will be discussed in Sec.~\ref{b3im}.

\subsection{\label{sec:path}ASCC reaction path for $\alpha$+$\alpha$$\leftrightarrow$$^{8}$Be  }

The numerical application of the ASCC method to determine a collective path for
the nuclear fusion or fission reactions demands a substantial computational cost.
Here, we present the result for the reaction path of $\alpha$+$\alpha$$\leftrightarrow$$^{8}$Be,
as the simplest example.
It can be regarded as either the fusion path of two
alpha particles or the fission path of $^{8}$Be.
The model space is the three-dimensional grid space of the rectangular box
of size $10\times10\times18$ fm$^{3}$ with mesh size equal to 1.0 fm.
The standard BKN energy density functional is adopted.

\begin{figure}
\begin{centering}
\includegraphics[height=5.5cm]{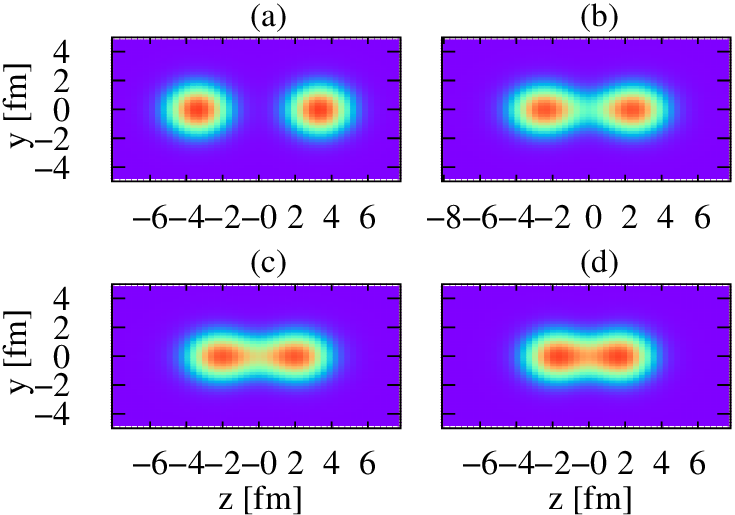}\quad\quad\includegraphics[height=5.5cm]{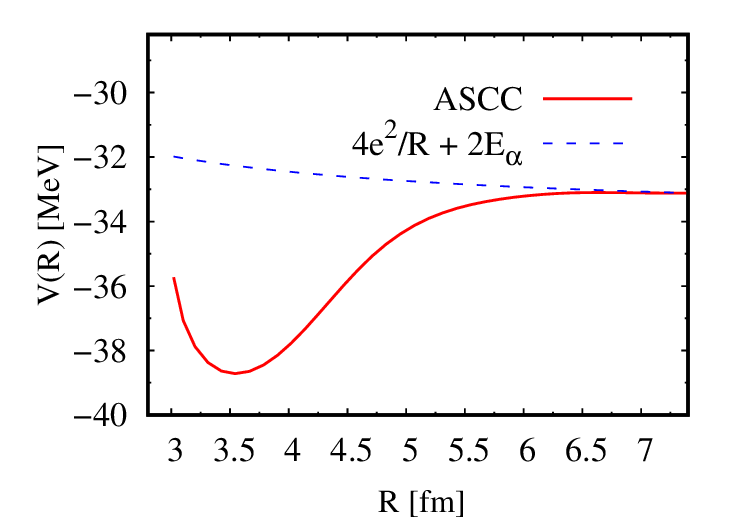}
\caption{Left panel:
Calculated density distributions of four points on
the ASCC collective fusion path $\alpha$+$\alpha$$\rightarrow$$^{8}$Be.
Inset (a) shows the density distribution of two well separated alpha particles at $R = 6.90$ fm,
inset (d) is the ground state of $^{8}$Be at $R = 3.55$ fm. Inset (b), (c) show the density
distributions at $R = 4.10$ fm, $5.40$ fm, respectively. Those on the $y-z$ plane
are plotted.
Right panel:
Potential energy as a function of $R$ shown by the red curve.
The blue the dashed line is calculated as $4e^{2}/R+2E_\alpha$ for reference.
}
\label{fig:density_and_potential}
\end{centering}
\end{figure}

Starting from the two ground states of $\alpha$ paticle and carrying out the iterative procedure
presented in Sec. \ref{sec:theo2}, we obtain a fusion path that
connects the two well separated alpha particles to the ground state of $^{8}$Be.
If we start the calculation from the ground state of $^{8}$Be,
the same reaction path, that represents fission of $^{8}$Be, can be obtained.
In the left four panels of Fig. \ref{fig:density_and_potential},
we show the calculated density distribution of four different points on
the obtained collective fusion path. The panel (a) shows the density distribution
of two alpha particles at $R = 6.90$ fm, (d) shows that
of the ground state of $^{8}$Be which corresponds to $R = 3.55$ fm.
Those of (b) and (c) show those
at $R = 5.40$ fm and $4.10$ fm, respectively.
The collective path smoothly evolves the separated two alpha particles
into the ground state of $^{8}$Be.

The right panel of Fig. \ref{fig:density_and_potential} shows the
potential energy along this collective path, as a function of $R$.
The dashed cure shows the point Coulomb potential, $4e/R+2E_{\rm \alpha}$,
with the ground state energy of a single alpha particle $E_{\rm \alpha}$.
With the BKN energy density functional, the $^{8}$Be is bound in the mean-field level.
The ground state of $^{8}$Be is located in the
potential minimum at $R = 3.55$ fm, while
the Coulomb barrier top is at $R = 6.50$ fm.
This ASCC collective path is self-consistently
generated by the iterative procedure presented in Sec.~\ref{sec:theo2}.
The generators $(\hat{Q}(q),\hat{P}(q))$ for the relative motion
are microscopically given.
Since the structure of the $^8$Be nucleus is very simple,
this potential surface is actually similar to that
of the constraint Hartree-Fock calculation.

%
\subsection{Inertial mass for $\alpha$+$\alpha$$\leftrightarrow$$^{8}$Be }
\label{sec:mass}

Upon the collective reaction path obtained,
the inertial mass with respect to the relative distance $R$, $M_{\rm ASCC}(R)$,
is calculated using Eq. (\ref{mass}).
In the asymptotic region, we expect the inertial
mass to be identical to the reduced mass,
$\mu_{\rm red}=A_{\rm pro}A_{\rm tar}m/(A_{\rm pro}+A_{\rm tar})$,
where $m$ is the nucleon mass.
For the current system $\alpha$+$\alpha$$\leftrightarrow$$^{8}$Be,
the value of $\mu_{\rm red}$ is expected to be $2m$.

The reduced mass $\mu_{\rm red}$ is justifiable when two alpha particles
are well separated.
However, it loses its validity as two particles approach each other.
A widely used approach to calculate inertial mass for
nuclear collective motion is the
``Constrained-Hartree-Fock-plus-cranking'' (CHF+cranking) approach \citep{Bar11}.
In this approach, the collective path is produced by the CHF calculation with
a constraining operator $\hat{O}$ given by hand, and the inertial mass is
calculated based on the cranking formula with respect to these CHF states.
The formula for the cranking mass can be derived
by the adiabatic perturbation \citep{RS80}.
In the present case of the one-dimensional motion,
based on the states constructed by the CHF calculation with
a given constraining operator $\hat{O}$,
the cranking formula reads \citep{Bar11}
\begin{equation}
M_{\rm cr}^{\rm NP}(R)=2 \sum_{n\in p,j\in h}
\frac{|\langle\varphi_n(R)|\partial/\partial R|\varphi_j(R)\rangle|^2}
{e_n(R)-e_j(R)} ,
\label{NP_cranking}
\end{equation}
where the single-particle states $\varphi_\mu$ and their energies $e_\mu$ are defined with
respect to $h_{\rm CHF}(\lambda)=h_{\rm HF}[\rho]-\lambda \hat{O}$,
\begin{equation}
h_{\rm CHF}(\lambda)|\varphi_\mu(\lambda)\rangle=e_\mu(\lambda))|\varphi_\mu(\lambda)\rangle ,
\quad \mu \in p, h .
\end{equation}
We may use any operator $\hat{O}$ as a constraint, as far as it can generate the states
with all the necessary values of $R=\langle\hat{R}\rangle$.
However, obviously the inertial mass $M(R)$ depends on this choice,
which is one of drawbacks of the CHF+cranking approach.

\begin{figure}
\begin{centering}
\includegraphics[width=0.90\columnwidth]{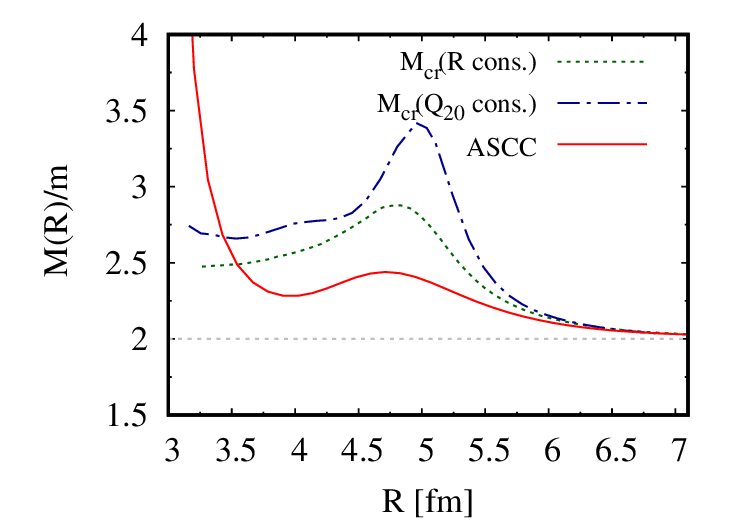}
\par\end{centering}
\caption{\label{fig:mass_cranking}(Color online)
Inertia masses $M_{\rm R}$ for the reaction
$\alpha$+$\alpha$$\leftrightarrow$$^{8}$Be as a function of relative distance $R$.
The solid (red) curve indicates the result of ASCC.
The other curves show the cranking masses of Eq. (\ref{NP_cranking})
calculated based on CHF states.
The dotted (green) and dash-dotted (blue) lines indicate the results
with constraints on $\hat{R}$ and $\hat{Q}_{20}$, respectively.
}
\end{figure}

In most of the reaction models,
the inertial mass with respect to $R$ is assumed to be
a constant value of $\mu_{\rm red}$.
Our study reveales how the inertia changes as a function of $R$.
In Fig. \ref{fig:mass_cranking},
both the ASCC and the cranking masses are presented.
For the cranking mass, since the CHF state needs to be prepared first.
We calculate the CHF states in two ways with different constraining operators $\hat{O}$;
the mass quadrupole operator $\hat{Q}_{20}$ and the relative distance $\hat{R}$ operator
of Eq.~(\ref{defr}).
The model space for both calculations are the same.
As we can see from figure \ref{fig:mass_cranking}, at large distance,
both methods asymptotically reproduce the reduced mass of $2m$, which is
the exact value for the relative motion between two alpha particles.
In the interior region where the two nuclei have merged into one system,
these three masses give very different values.
Generally the cranking mass is found to be larger than the
ASCC mass, especially at around $R = 4.7$ fm where all the
three masses develop a bump structure.

The difference between the ASCC and the cranking masses attributes to
several factors.
One is due to the fact that the cranking formula neglects residual fields
induced by the density fluctuation.
Another is that the constraining operators affect the single-particle energies
$e_\mu(R)$.
We also note that the cranking masses
obtained with different constraints give very different values.
This is true even at the HF ground state ($R=3.55$ fm), in which
the single-particle states $\ket{\varphi_\mu(R)}$ and their single-particle energies
$e_\mu(R)$ are all identical to each other.
This is because the derivative $\partial/\partial R$ gives different values,
since the different constraint produces different states away from the HF ground state.
This ambiguity exposes another drawback of the CHF+cranking approach,
while the ASCC mass has an advantage
that the collective coordinate as well as the wave functions
are self-consistently calculated rather than artificially assumed.

\subsection{\label{b3im}Impact of time-odd potential}
All the results shown so far are obtained with the standard BKN energy density functional
that has no derivative terms.
Therefore, the nucleon's effective mass
is identical to the bare nucleon mass. However, most of realistic effective
interactions have effective mass smaller than the bare
mass, typically $m^{*}/m \sim 0.7$. In such cases, an
improper treatment of the collective dynamics leads
to a wrong answer for the collective inertial mass \citep{TV62}.
This change in the effective mass
typically comes from the term $\rho\tau$ in the Skyrme
energy density functional, which should accompany the term $-\mathbf{j}^2$
to restore the Galilean symmetry \citep{TV62,BM75}. 
These terms are absent in the standard BKN functional.

To investigate the effect of the time-odd mean-field potential on the collective inertial mass,
we add the term $B_{3}(\rho\tau-\textbf{j}^{2})$ to the original BKN energy density functional.
The modified BKN energy density functional reads,
\begin{eqnarray}
	E[\rho] &=& \int \frac{1}{2m} \tau(\mathbf{r}) d\mathbf{r}
	+\int d\mathbf{r} \left\{
            \frac{3}{8}t_{0}\rho^2(\mathbf{r})
           + \frac{1}{16}t_{3}\rho^{3}(\mathbf{r})\right\}
              \nonumber\\
   &&+\int \int d\mathbf{r} d\mathbf{r}' \rho(\mathbf{r})v(\mathbf{r}-\mathbf{r}')\rho(\mathbf{r}')
                 \nonumber\\
	&&+B_{3} \int d\mathbf{r}
	\left\{ \rho(\mathbf{r})\tau(\mathbf{r}) - \mathbf{j}^{2}(\mathbf{r})\right\}
           \label{BKND}
\end{eqnarray}
where $\rho(\mathbf{r})$, $\tau(\mathbf{r})$, and $\mathbf{j}(\mathbf{r})$
are the isoscalar density, the isoscalar kinetic density, and the isoscalar
current density, respectively. 
In equation (\ref{BKND}), $v(\vec{r})$ is the sum of the Yukawa and the Coulomb potentials
\citep{BKN76}.
%
The variation of the total energy with respect to the density
(or equivalently single-particle wave functions)
defines the single-particle (Hartree-Fock) Hamiltonian.
In the present case, the single-particle Hamiltonian turns out to be
\begin{eqnarray}
h[\rho] &=& -\nabla \frac{1}{2m^{*}(\mathbf{r})}\nabla
           + \frac{3}{4}t_{0}\rho(\mathbf{r})
           + \frac{3}{16}t_{3}\rho^{2}(\mathbf{r})
           + \int d\mathbf{r'} v(\mathbf{r}-\mathbf{r'})\rho(\mathbf{r'}), \nonumber\\
	   &&+B_{3}(\tau(\mathbf{r})+i\nabla\cdot \mathbf{j}(\mathbf{r}))
	   +2iB_{3}\mathbf{j}(\mathbf{r})\cdot\nabla
\end{eqnarray}
where the effective mass is now deviated from bare nucleon mass
\begin{eqnarray}
	\frac{\hbar^{2}}{2m^{*}(\mathbf{r})}=\frac{\hbar^{2}}{2m}+B_{3}\rho(\mathbf{r}).
\end{eqnarray}
For the time-even states, such as the ground state of even-even nuclei,
the current density disappears, $\mathbf{j}=0$.
Even though, these terms play an important role in the collective inertial mass.
The parameter $B_{3}\neq 0$ provides the effective mass and the time-odd effect.
The rest of the parameters are the same as those in reference \citep{BKN76}.

\begin{figure}
\begin{centering}
\includegraphics[height=5.5cm]{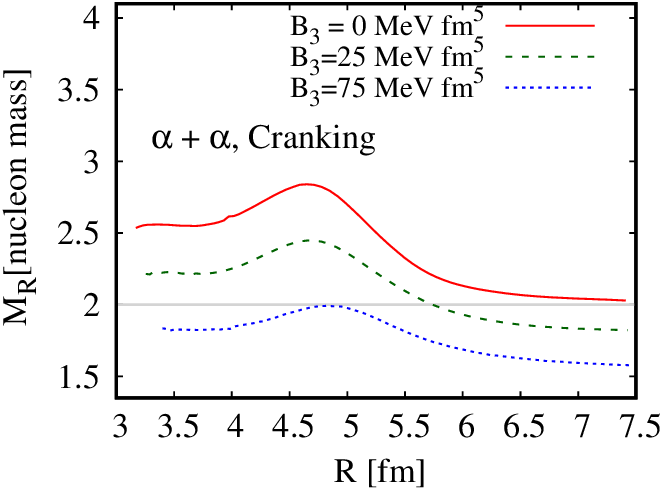}\quad\includegraphics[height=5.5cm]{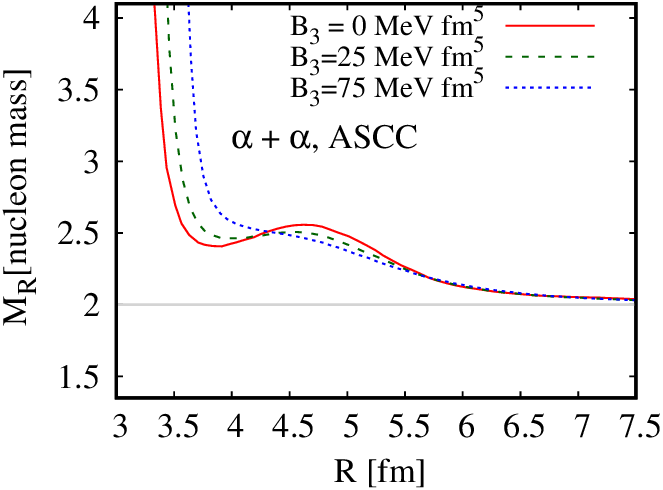}
\caption{
Relative inertial masses in the presence of time-odd mean-field potential for the
reaction $\alpha$+$\alpha$$\leftrightarrow$$^{8}$Be as a function of relative distance $R$.
The results of the cranking masses are shown in the left panel and those of the ASCC method
are shown in the right panel.
The solid (red), dashed (green) and dotted (blue) curves show the results calculated
with $B_{3}= 0, 25$, and 75 MeV fm$^{5}$ respectively.
}
\label{fig:time-odd}
\end{centering}
\end{figure}

To examine the impact of the time-odd terms on the inertial mass,
in Fig. (\ref{fig:time-odd}) we show
$M(R)$ calculated with and without the $B_3$ term.
When the time-odd terms are absent, $B3 = 0$,
both the ASCC and the cranking formula reproduce the
$\alpha + \alpha$ reduced mass in the asymptotic limit ($R\rightarrow\infty$).
However, the cranking formula fails to do so
with $B3 \neq 0$. As the value of $B_{3}$ increases,
the asymptotic cranking mass decreases.
This can be naively expected from the reduction of the effective mass from the bare mass.
In contrast, the ASCC
inertial mass converges to the correct reduced mass, no
matter what $B_{3}$ values are. This means that the
ASCC method is capable of taking into account the time-odd effect
and recovering the exact Galilean symmetry.

Another inertial mass indispensable in the collective Hamiltonian of nuclear reaction
models is the rotational moments of inertia.
The rotational motion is a Nambu-Goldstone (NG) mode.
To calculate this, we utilize a method proposed in the reference \citep{Hino2015},
where the inertial masses of the NG modes are calculated from
the zero-frequency linear response with the momentum operator of the NG modes.
The formulation has been tested in the cases of translational
and pairing rotational modes, showing high precision and efficiency.
Based on the collective path obtained,
we apply this technique to calculate the rotational moments of inertia.
\begin{figure}
\begin{centering}
\includegraphics[height=5.5cm]{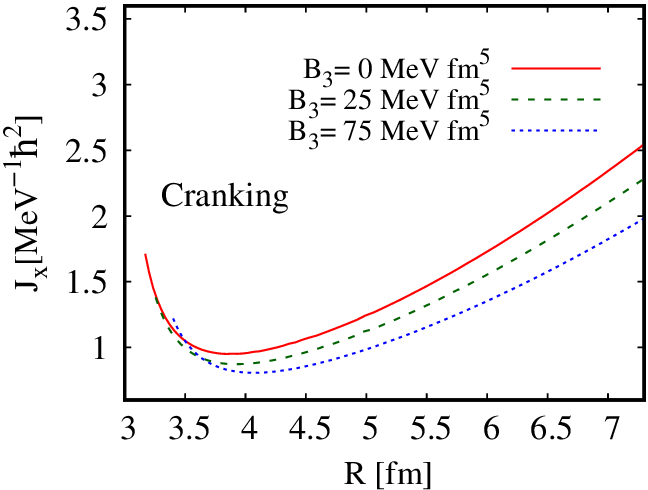}\quad\includegraphics[height=5.5cm]{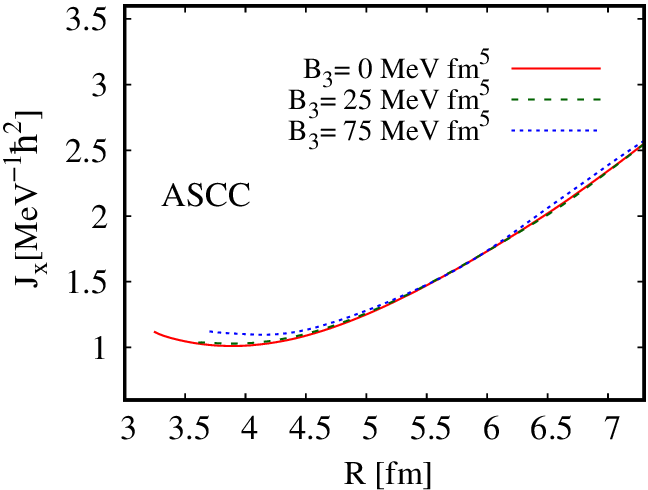}
\caption{
Rotational moments of inertias in the presence of time-odd mean-field potential for the
system $\alpha$+$\alpha$ as a function of relative distance $R$.
The results of cranking formula are shown in the left panel and the results of ASCC
are shown in the right panel.
The solid(red), dashed(green) and dotted(blue) curves show the results calculated
with $B_{3}= 0, 25, 75$ MeV fm$^{5}$ respectively as labeled in the figure.
}
\label{fig:MoI}
\end{centering}
\end{figure}
%

In figure \ref{fig:MoI}, the calculated moments of inertias are presented.
With $B_{3} = 0$,
the moments of inertia calculated with the ASCC and with the cranking formula
well agree with each other in the asymptotic region of large $R$.
The value is equal to the point-mass approximation in which the point $\alpha$ particles
are assumed at the center of mass of each $\alpha$ particle.
However, when non-zero $B_{3}$ comes in, the cranking mass
formula can no longer reproduce this asymptotic value.
Similar to the case of relative motion, as the value of $B_{3}$ increases,
the asymptotic moments of inertia decrease and deviate from the asymptotic value.
In contrast, the ASCC method provides the moments of inertia almost invariant
with respect to the $B_3$ values.
These results show again that, compared with the cranking formula,
the ASCC method gives the collective inertial masses by properly
taking into account the time-odd effects.

\section{Summary and discussion}
\label{sec:sum}
Based on the ASCC theory, we presented a method to determine the
collective reaction path for the nuclear reaction as the large amplitude collective motion.
This method is applied to the fusion/fission $\alpha$+$\alpha$$\leftrightarrow$$^{8}$Be,
using the BKN energy density functional.
In the three-dimensional coordinate-space representation, the reaction path,
the collective potential, as well as the inertial masses are self-consistently calculated.
We compare the ASCC results with those of the CHF+cranking method.
Since the reaction system is very simple,
there is no significant difference between
the calculated CHF reaction paths with different constraint operators.
Despite of this similarity in the CHF states,
the inertial masses calculated with the cranking formula turn out to sensitively depend on the
choice of the constraint operator. 
The ASCC method is able to remove this ambiguity in the inertial mass,
by taking into account the residual effects
caused by the density fluctuation. 

We add a term,
which introduce the effective mass and time-odd mean fields,
to the standard BKN energy density functional,
to examine the effect of these terms on
the inertial masses for both the relative
and rotational motions.
In the presence of time-odd term,
the cranking formula fails to preserve the correct asymptotic values,
while the validity of ASCC mass is not affected by the introduction of the effective mass.
The time-odd mean-fields properly recover the Galilean symmetry, leading to the exact values of
the asymptotic inertial mass.
This is found to be true in both relative and rotational motions. 
With this property, we are quite confident that the ASCC method is promising
to be applied to the modern nuclear energy density functionals,
and make advanced microscopic theoretical analysis on various nuclear reaction models.
Another important issue is the inclusion of the paring correlation,
which may influence not only static
but also dynamical nuclear properties.
In order to keep the lowest-energy configuration during the collective motion,
the pairing interaction is known to play a key role \citep{NMMS16}.
Therefore, we may expect significant impact on
both the collective inertial masses and the reaction paths.
To study the above issues are our future tasks.

\section*{Acknowledgments}

This work is supported in part by JSPS KAKENHI Grant No. 19H05142
and No. 18H01209,
and also by JSPS-NSFC Bilateral Program for Joint Research Project on
Nuclear mass and life for unravelling mysteries of r-process.
This research in part used computational resources provided
through the HPCI Sysytem Research Project (Project ID: hp190031)
and by Multidisciplinary Cooperative Research Program
in Center for Computational Sciences, University of Tsukuba.

\begin{acknowledgments}
This work is supported in part by JSPS KAKENHI Grant No. 19H05142, and also by JSPS-NSFC
Bilateral Program for Joint Research Project on Nuclear mass and life for unravelling mysteries of r-
process. This research in part used computational resources provided through the HPCI Sysytem Research
Project (Project ID: hp190031) and by Multidisciplinary Cooperative Research Program in Center for
Computational Sciences, University of Tsukuba.
This work is supported in part by JSPS KAKENHI Grants No. 25287065
and by Interdisciplinary Computational Science Program in CCS,
University of Tsukuba.
\end{acknowledgments}

\bibliographystyle{apsrev4-1}
\bibliography{./test}

\begin{thebibliography}{30}%
\makeatletter
\providecommand \@ifxundefined [1]{%
 \@ifx{#1\undefined}
}%
\providecommand \@ifnum [1]{%
 \ifnum #1\expandafter \@firstoftwo
 \else \expandafter \@secondoftwo
 \fi
}%
\providecommand \@ifx [1]{%
 \ifx #1\expandafter \@firstoftwo
 \else \expandafter \@secondoftwo
 \fi
}%
\providecommand \natexlab [1]{#1}%
\providecommand \enquote  [1]{``#1''}%
\providecommand \bibnamefont  [1]{#1}%
\providecommand \bibfnamefont [1]{#1}%
\providecommand \citenamefont [1]{#1}%
\providecommand \href@noop [0]{\@secondoftwo}%
\providecommand \href [0]{\begingroup \@sanitize@url \@href}%
\providecommand \@href[1]{\@@startlink{#1}\@@href}%
\providecommand \@@href[1]{\endgroup#1\@@endlink}%
\providecommand \@sanitize@url [0]{\catcode `\\12\catcode `\$12\catcode
  `\&12\catcode `\#12\catcode `\^12\catcode `\_12\catcode `\%12\relax}%
\providecommand \@@startlink[1]{}%
\providecommand \@@endlink[0]{}%
\providecommand \url  [0]{\begingroup\@sanitize@url \@url }%
\providecommand \@url [1]{\endgroup\@href {#1}{\urlprefix }}%
\providecommand \urlprefix  [0]{URL }%
\providecommand \Eprint [0]{\href }%
\providecommand \doibase [0]{http://dx.doi.org/}%
\providecommand \selectlanguage [0]{\@gobble}%
\providecommand \bibinfo  [0]{\@secondoftwo}%
\providecommand \bibfield  [0]{\@secondoftwo}%
\providecommand \translation [1]{[#1]}%
\providecommand \BibitemOpen [0]{}%
\providecommand \bibitemStop [0]{}%
\providecommand \bibitemNoStop [0]{.\EOS\space}%
\providecommand \EOS [0]{\spacefactor3000\relax}%
\providecommand \BibitemShut  [1]{\csname bibitem#1\endcsname}%
\let\auto@bib@innerbib\@empty
\bibitem [{\citenamefont {Negele}(1982)}]{Neg82}%
  \BibitemOpen
  \bibfield  {author} {\bibinfo {author} {\bibfnamefont {J.~W.}\ \bibnamefont
  {Negele}},\ }\href {\doibase 10.1103/RevModPhys.54.913} {\bibfield  {journal}
  {\bibinfo  {journal} {Rev. Mod. Phys.}\ }\textbf {\bibinfo {volume} {54}},\
  \bibinfo {pages} {913} (\bibinfo {year} {1982})}\BibitemShut {NoStop}%
\bibitem [{\citenamefont {Simenel}(2012)}]{Sim12}%
  \BibitemOpen
  \bibfield  {author} {\bibinfo {author} {\bibfnamefont {C.}~\bibnamefont
  {Simenel}},\ }\href {\doibase 10.1140/epja/i2012-12152-0} {\bibfield
  {journal} {\bibinfo  {journal} {The European Physical Journal A}\ }\textbf
  {\bibinfo {volume} {48}},\ \bibinfo {pages} {1} (\bibinfo {year}
  {2012})}\BibitemShut {NoStop}%
\bibitem [{\citenamefont {Nakatsukasa}(2012)}]{Nak12}%
  \BibitemOpen
  \bibfield  {author} {\bibinfo {author} {\bibfnamefont {T.}~\bibnamefont
  {Nakatsukasa}},\ }\href {\doibase 10.1093/ptep/pts016} {\bibfield  {journal}
  {\bibinfo  {journal} {Prog. Theor. Exp. Phys.}\ }\textbf {\bibinfo {volume}
  {2012}} (\bibinfo {year} {2012}),\ 10.1093/ptep/pts016},\ \bibinfo {note}
  {01A207}\BibitemShut {NoStop}%
\bibitem [{\citenamefont {Maruhn}\ \emph {et~al.}(2014)\citenamefont {Maruhn},
  \citenamefont {Reinhard}, \citenamefont {Stevenson},\ and\ \citenamefont
  {Umar}}]{MRSU14}%
  \BibitemOpen
  \bibfield  {author} {\bibinfo {author} {\bibfnamefont {J.~A.}\ \bibnamefont
  {Maruhn}}, \bibinfo {author} {\bibfnamefont {P.-G.}\ \bibnamefont
  {Reinhard}}, \bibinfo {author} {\bibfnamefont {P.~D.}\ \bibnamefont
  {Stevenson}}, \ and\ \bibinfo {author} {\bibfnamefont {A.~S.}\ \bibnamefont
  {Umar}},\ }\href {\doibase 10.1016/j.cpc.2014.04.008} {\bibfield  {journal}
  {\bibinfo  {journal} {Computer Physics Communications}\ }\textbf {\bibinfo
  {volume} {185}},\ \bibinfo {pages} {2195} (\bibinfo {year}
  {2014})}\BibitemShut {NoStop}%
\bibitem [{\citenamefont {Nakatsukasa}\ \emph
  {et~al.}(2016{\natexlab{a}})\citenamefont {Nakatsukasa}, \citenamefont
  {Matsuyanagi}, \citenamefont {Matsuo},\ and\ \citenamefont
  {Yabana}}]{NMMY16}%
  \BibitemOpen
  \bibfield  {author} {\bibinfo {author} {\bibfnamefont {T.}~\bibnamefont
  {Nakatsukasa}}, \bibinfo {author} {\bibfnamefont {K.}~\bibnamefont
  {Matsuyanagi}}, \bibinfo {author} {\bibfnamefont {M.}~\bibnamefont {Matsuo}},
  \ and\ \bibinfo {author} {\bibfnamefont {K.}~\bibnamefont {Yabana}},\ }\href
  {\doibase 10.1103/RevModPhys.88.045004} {\bibfield  {journal} {\bibinfo
  {journal} {Rev. Mod. Phys.}\ }\textbf {\bibinfo {volume} {88}},\ \bibinfo
  {pages} {045004} (\bibinfo {year} {2016}{\natexlab{a}})}\BibitemShut
  {NoStop}%
\bibitem [{\citenamefont {Ring}\ and\ \citenamefont {Schuck}(1980)}]{RS80}%
  \BibitemOpen
  \bibfield  {author} {\bibinfo {author} {\bibfnamefont {P.}~\bibnamefont
  {Ring}}\ and\ \bibinfo {author} {\bibfnamefont {P.}~\bibnamefont {Schuck}},\
  }\href@noop {} {\emph {\bibinfo {title} {The Nuclear Many-Body Problem}}}\
  (\bibinfo  {publisher} {Springer-Verlag, New York},\ \bibinfo {year}
  {1980})\BibitemShut {NoStop}%
\bibitem [{\citenamefont {Blaizot}\ and\ \citenamefont {Ripka}(1986)}]{BR86}%
  \BibitemOpen
  \bibfield  {author} {\bibinfo {author} {\bibfnamefont {J.~P.}\ \bibnamefont
  {Blaizot}}\ and\ \bibinfo {author} {\bibfnamefont {G.}~\bibnamefont
  {Ripka}},\ }\href {\doibase 10.1080/00107518608211031} {\emph {\bibinfo
  {title} {Quantum theory of finite systems}}}\ (\bibinfo  {publisher} {MIT
  Press, Cambridge},\ \bibinfo {year} {1986})\BibitemShut {NoStop}%
\bibitem [{\citenamefont {Baranger}\ and\ \citenamefont
  {Kumar}(1968{\natexlab{a}})}]{BK68-1}%
  \BibitemOpen
  \bibfield  {author} {\bibinfo {author} {\bibfnamefont {M.}~\bibnamefont
  {Baranger}}\ and\ \bibinfo {author} {\bibfnamefont {K.}~\bibnamefont
  {Kumar}},\ }\href {\doibase 10.1016/0375-9474(68)90371-0} {\bibfield
  {journal} {\bibinfo  {journal} {Nuclear Physics A}\ }\textbf {\bibinfo
  {volume} {110}},\ \bibinfo {pages} {490} (\bibinfo {year}
  {1968}{\natexlab{a}})}\BibitemShut {NoStop}%
\bibitem [{\citenamefont {Baranger}\ and\ \citenamefont
  {Kumar}(1968{\natexlab{b}})}]{BK68-2}%
  \BibitemOpen
  \bibfield  {author} {\bibinfo {author} {\bibfnamefont {M.}~\bibnamefont
  {Baranger}}\ and\ \bibinfo {author} {\bibfnamefont {K.}~\bibnamefont
  {Kumar}},\ }\href {\doibase 10.1016/0375-9474(68)90370-9} {\bibfield
  {journal} {\bibinfo  {journal} {Nuclear Physics A}\ }\textbf {\bibinfo
  {volume} {122}},\ \bibinfo {pages} {241} (\bibinfo {year}
  {1968}{\natexlab{b}})}\BibitemShut {NoStop}%
\bibitem [{\citenamefont {Yuldashbaeva}\ \emph {et~al.}(1999)\citenamefont
  {Yuldashbaeva}, \citenamefont {Libert}, \citenamefont {Quentin},\ and\
  \citenamefont {Girod}}]{YLQG99}%
  \BibitemOpen
  \bibfield  {author} {\bibinfo {author} {\bibfnamefont {E.}~\bibnamefont
  {Yuldashbaeva}}, \bibinfo {author} {\bibfnamefont {J.}~\bibnamefont
  {Libert}}, \bibinfo {author} {\bibfnamefont {P.}~\bibnamefont {Quentin}}, \
  and\ \bibinfo {author} {\bibfnamefont {M.}~\bibnamefont {Girod}},\ }\href
  {\doibase 10.1016/S0370-2693(99)00836-9.} {\bibfield  {journal} {\bibinfo
  {journal} {Physics Letters B}\ }\textbf {\bibinfo {volume} {461}},\ \bibinfo
  {pages} {1} (\bibinfo {year} {1999})}\BibitemShut {NoStop}%
\bibitem [{\citenamefont {Marumori}\ \emph {et~al.}(1980)\citenamefont
  {Marumori}, \citenamefont {Maskawa}, \citenamefont {Sakata},\ and\
  \citenamefont {Kuriyama}}]{MMSK80}%
  \BibitemOpen
  \bibfield  {author} {\bibinfo {author} {\bibfnamefont {T.}~\bibnamefont
  {Marumori}}, \bibinfo {author} {\bibfnamefont {T.}~\bibnamefont {Maskawa}},
  \bibinfo {author} {\bibfnamefont {F.}~\bibnamefont {Sakata}}, \ and\ \bibinfo
  {author} {\bibfnamefont {A.}~\bibnamefont {Kuriyama}},\ }\href {\doibase
  10.1143/PTP.64.1294} {\bibfield  {journal} {\bibinfo  {journal} {Prog. Theor.
  Phys.}\ }\textbf {\bibinfo {volume} {64}},\ \bibinfo {pages} {1294} (\bibinfo
  {year} {1980})}\BibitemShut {NoStop}%
\bibitem [{\citenamefont {Matsuo}\ \emph {et~al.}(2000)\citenamefont {Matsuo},
  \citenamefont {Nakatsukasa},\ and\ \citenamefont {Matsuyanagi}}]{MNM00}%
  \BibitemOpen
  \bibfield  {author} {\bibinfo {author} {\bibfnamefont {M.}~\bibnamefont
  {Matsuo}}, \bibinfo {author} {\bibfnamefont {T.}~\bibnamefont {Nakatsukasa}},
  \ and\ \bibinfo {author} {\bibfnamefont {K.}~\bibnamefont {Matsuyanagi}},\
  }\href {\doibase 10.1143/PTP.103.959} {\bibfield  {journal} {\bibinfo
  {journal} {Prog. Theor. Phys.}\ }\textbf {\bibinfo {volume} {103 (5)}},\
  \bibinfo {pages} {959} (\bibinfo {year} {2000})}\BibitemShut {NoStop}%
\bibitem [{\citenamefont {Hinohara}\ \emph {et~al.}(2007)\citenamefont
  {Hinohara}, \citenamefont {Nakatsukasa}, \citenamefont {Matsuo},\ and\
  \citenamefont {Matsuyanagi}}]{HNMM07}%
  \BibitemOpen
  \bibfield  {author} {\bibinfo {author} {\bibfnamefont {N.}~\bibnamefont
  {Hinohara}}, \bibinfo {author} {\bibfnamefont {T.}~\bibnamefont
  {Nakatsukasa}}, \bibinfo {author} {\bibfnamefont {M.}~\bibnamefont {Matsuo}},
  \ and\ \bibinfo {author} {\bibfnamefont {K.}~\bibnamefont {Matsuyanagi}},\
  }\href {\doibase 10.1143/PTP.117.451} {\bibfield  {journal} {\bibinfo
  {journal} {Prog. Theor. Phys.}\ }\textbf {\bibinfo {volume} {117 (3)}},\
  \bibinfo {pages} {451} (\bibinfo {year} {2007})}\BibitemShut {NoStop}%
\bibitem [{\citenamefont {Hinohara}\ \emph {et~al.}(2009)\citenamefont
  {Hinohara}, \citenamefont {Nakatsukasa}, \citenamefont {Matsuo},\ and\
  \citenamefont {Matsuyanagi}}]{HNMM09}%
  \BibitemOpen
  \bibfield  {author} {\bibinfo {author} {\bibfnamefont {N.}~\bibnamefont
  {Hinohara}}, \bibinfo {author} {\bibfnamefont {T.}~\bibnamefont
  {Nakatsukasa}}, \bibinfo {author} {\bibfnamefont {M.}~\bibnamefont {Matsuo}},
  \ and\ \bibinfo {author} {\bibfnamefont {K.}~\bibnamefont {Matsuyanagi}},\
  }\href {\doibase 10.1103/PhysRevC.80.014305} {\bibfield  {journal} {\bibinfo
  {journal} {Phys. Rev. C}\ }\textbf {\bibinfo {volume} {80}},\ \bibinfo
  {pages} {014305} (\bibinfo {year} {2009})}\BibitemShut {NoStop}%
\bibitem [{\citenamefont {Hinohara}\ \emph {et~al.}(2012)\citenamefont
  {Hinohara}, \citenamefont {Li}, \citenamefont {Nakatsukasa}, \citenamefont
  {Nik\ifmmode \check{s}\else \v{s}\fi{}i\ifmmode~\acute{c}\else \'{c}\fi{}},\
  and\ \citenamefont {Vretenar}}]{HLNNV12}%
  \BibitemOpen
  \bibfield  {author} {\bibinfo {author} {\bibfnamefont {N.}~\bibnamefont
  {Hinohara}}, \bibinfo {author} {\bibfnamefont {Z.~P.}\ \bibnamefont {Li}},
  \bibinfo {author} {\bibfnamefont {T.}~\bibnamefont {Nakatsukasa}}, \bibinfo
  {author} {\bibfnamefont {T.}~\bibnamefont {Nik\ifmmode \check{s}\else
  \v{s}\fi{}i\ifmmode~\acute{c}\else \'{c}\fi{}}}, \ and\ \bibinfo {author}
  {\bibfnamefont {D.}~\bibnamefont {Vretenar}},\ }\href {\doibase
  10.1103/PhysRevC.85.024323} {\bibfield  {journal} {\bibinfo  {journal} {Phys.
  Rev. C}\ }\textbf {\bibinfo {volume} {85}},\ \bibinfo {pages} {024323}
  (\bibinfo {year} {2012})}\BibitemShut {NoStop}%
\bibitem [{\citenamefont {Sato}\ \emph {et~al.}(2012)\citenamefont {Sato},
  \citenamefont {Hinohara}, \citenamefont {Yoshida}, \citenamefont
  {Nakatsukasa}, \citenamefont {Matsuo},\ and\ \citenamefont
  {Matsuyanagi}}]{SHYNMM12}%
  \BibitemOpen
  \bibfield  {author} {\bibinfo {author} {\bibfnamefont {K.}~\bibnamefont
  {Sato}}, \bibinfo {author} {\bibfnamefont {N.}~\bibnamefont {Hinohara}},
  \bibinfo {author} {\bibfnamefont {K.}~\bibnamefont {Yoshida}}, \bibinfo
  {author} {\bibfnamefont {T.}~\bibnamefont {Nakatsukasa}}, \bibinfo {author}
  {\bibfnamefont {M.}~\bibnamefont {Matsuo}}, \ and\ \bibinfo {author}
  {\bibfnamefont {K.}~\bibnamefont {Matsuyanagi}},\ }\href {\doibase
  10.1103/PhysRevC.86.024316} {\bibfield  {journal} {\bibinfo  {journal} {Phys.
  Rev. C}\ }\textbf {\bibinfo {volume} {86}},\ \bibinfo {pages} {024316}
  (\bibinfo {year} {2012})}\BibitemShut {NoStop}%
\bibitem [{\citenamefont {Davies}\ \emph {et~al.}(1980)\citenamefont {Davies},
  \citenamefont {Flocard}, \citenamefont {Krieger},\ and\ \citenamefont
  {Weiss}}]{DFKW80}%
  \BibitemOpen
  \bibfield  {author} {\bibinfo {author} {\bibfnamefont {K.}~\bibnamefont
  {Davies}}, \bibinfo {author} {\bibfnamefont {H.}~\bibnamefont {Flocard}},
  \bibinfo {author} {\bibfnamefont {S.}~\bibnamefont {Krieger}}, \ and\
  \bibinfo {author} {\bibfnamefont {M.}~\bibnamefont {Weiss}},\ }\href
  {\doibase 10.1016/0375-9474(80)90509-6} {\bibfield  {journal} {\bibinfo
  {journal} {Nuclear Physics A}\ }\textbf {\bibinfo {volume} {342}},\ \bibinfo
  {pages} {111} (\bibinfo {year} {1980})}\BibitemShut {NoStop}%
\bibitem [{\citenamefont {Nakatsukasa}\ \emph {et~al.}(2007)\citenamefont
  {Nakatsukasa}, \citenamefont {Inakura},\ and\ \citenamefont
  {Yabana}}]{NIY07}%
  \BibitemOpen
  \bibfield  {author} {\bibinfo {author} {\bibfnamefont {T.}~\bibnamefont
  {Nakatsukasa}}, \bibinfo {author} {\bibfnamefont {T.}~\bibnamefont
  {Inakura}}, \ and\ \bibinfo {author} {\bibfnamefont {K.}~\bibnamefont
  {Yabana}},\ }\href {\doibase 10.1103/PhysRevC.76.024318} {\bibfield
  {journal} {\bibinfo  {journal} {Phys. Rev. C}\ }\textbf {\bibinfo {volume}
  {76}},\ \bibinfo {pages} {024318} (\bibinfo {year} {2007})}\BibitemShut
  {NoStop}%
\bibitem [{\citenamefont {Inakura}\ \emph {et~al.}(2009)\citenamefont
  {Inakura}, \citenamefont {Nakatsukasa},\ and\ \citenamefont
  {Yabana}}]{INY09}%
  \BibitemOpen
  \bibfield  {author} {\bibinfo {author} {\bibfnamefont {T.}~\bibnamefont
  {Inakura}}, \bibinfo {author} {\bibfnamefont {T.}~\bibnamefont
  {Nakatsukasa}}, \ and\ \bibinfo {author} {\bibfnamefont {K.}~\bibnamefont
  {Yabana}},\ }\href {\doibase 10.1103/PhysRevC.80.044301} {\bibfield
  {journal} {\bibinfo  {journal} {Phys. Rev. C}\ }\textbf {\bibinfo {volume}
  {80}},\ \bibinfo {pages} {044301} (\bibinfo {year} {2009})}\BibitemShut
  {NoStop}%
\bibitem [{\citenamefont {Avogadro}\ and\ \citenamefont
  {Nakatsukasa}(2011)}]{AN11}%
  \BibitemOpen
  \bibfield  {author} {\bibinfo {author} {\bibfnamefont {P.}~\bibnamefont
  {Avogadro}}\ and\ \bibinfo {author} {\bibfnamefont {T.}~\bibnamefont
  {Nakatsukasa}},\ }\href {\doibase 10.1103/PhysRevC.84.014314} {\bibfield
  {journal} {\bibinfo  {journal} {Phys. Rev. C}\ }\textbf {\bibinfo {volume}
  {84}},\ \bibinfo {pages} {014314} (\bibinfo {year} {2011})}\BibitemShut
  {NoStop}%
\bibitem [{\citenamefont {Avogadro}\ and\ \citenamefont
  {Nakatsukasa}(2013)}]{AN13}%
  \BibitemOpen
  \bibfield  {author} {\bibinfo {author} {\bibfnamefont {P.}~\bibnamefont
  {Avogadro}}\ and\ \bibinfo {author} {\bibfnamefont {T.}~\bibnamefont
  {Nakatsukasa}},\ }\href {\doibase 10.1103/PhysRevC.87.014331} {\bibfield
  {journal} {\bibinfo  {journal} {Phys. Rev. C}\ }\textbf {\bibinfo {volume}
  {87}},\ \bibinfo {pages} {014331} (\bibinfo {year} {2013})}\BibitemShut
  {NoStop}%
\bibitem [{\citenamefont {Wen}\ and\ \citenamefont {Nakatsukasa}(2016)}]{WN16}%
  \BibitemOpen
  \bibfield  {author} {\bibinfo {author} {\bibfnamefont {K.}~\bibnamefont
  {Wen}}\ and\ \bibinfo {author} {\bibfnamefont {T.}~\bibnamefont
  {Nakatsukasa}},\ }\href {\doibase 10.1103/PhysRevC.94.054618} {\bibfield
  {journal} {\bibinfo  {journal} {Phys. Rev. C}\ }\textbf {\bibinfo {volume}
  {94}},\ \bibinfo {pages} {054618} (\bibinfo {year} {2016})}\BibitemShut
  {NoStop}%
\bibitem [{\citenamefont {Egido}\ \emph {et~al.}(2016)\citenamefont {Egido},
  \citenamefont {Borrajo},\ and\ \citenamefont {Rodr\'{\i}guez}}]{EBR16}%
  \BibitemOpen
  \bibfield  {author} {\bibinfo {author} {\bibfnamefont {J.~L.}\ \bibnamefont
  {Egido}}, \bibinfo {author} {\bibfnamefont {M.}~\bibnamefont {Borrajo}}, \
  and\ \bibinfo {author} {\bibfnamefont {T.~R.}\ \bibnamefont
  {Rodr\'{\i}guez}},\ }\href {\doibase 10.1103/PhysRevLett.116.052502}
  {\bibfield  {journal} {\bibinfo  {journal} {Phys. Rev. Lett.}\ }\textbf
  {\bibinfo {volume} {116}},\ \bibinfo {pages} {052502} (\bibinfo {year}
  {2016})}\BibitemShut {NoStop}%
\bibitem [{\citenamefont {Li}\ \emph {et~al.}(2012)\citenamefont {Li},
  \citenamefont {Nik\ifmmode \check{s}\else \v{s}\fi{}i\ifmmode~\acute{c}\else
  \'{c}\fi{}}, \citenamefont {Ring}, \citenamefont {Vretenar}, \citenamefont
  {Yao},\ and\ \citenamefont {Meng}}]{Li12}%
  \BibitemOpen
  \bibfield  {author} {\bibinfo {author} {\bibfnamefont {Z.~P.}\ \bibnamefont
  {Li}}, \bibinfo {author} {\bibfnamefont {T.}~\bibnamefont {Nik\ifmmode
  \check{s}\else \v{s}\fi{}i\ifmmode~\acute{c}\else \'{c}\fi{}}}, \bibinfo
  {author} {\bibfnamefont {P.}~\bibnamefont {Ring}}, \bibinfo {author}
  {\bibfnamefont {D.}~\bibnamefont {Vretenar}}, \bibinfo {author}
  {\bibfnamefont {J.~M.}\ \bibnamefont {Yao}}, \ and\ \bibinfo {author}
  {\bibfnamefont {J.}~\bibnamefont {Meng}},\ }\href {\doibase
  10.1103/PhysRevC.86.034334} {\bibfield  {journal} {\bibinfo  {journal} {Phys.
  Rev. C}\ }\textbf {\bibinfo {volume} {86}},\ \bibinfo {pages} {034334}
  (\bibinfo {year} {2012})}\BibitemShut {NoStop}%
\bibitem [{\citenamefont {Bonche}\ \emph {et~al.}(1976)\citenamefont {Bonche},
  \citenamefont {Koonin},\ and\ \citenamefont {Negele}}]{BKN76}%
  \BibitemOpen
  \bibfield  {author} {\bibinfo {author} {\bibfnamefont {P.}~\bibnamefont
  {Bonche}}, \bibinfo {author} {\bibfnamefont {S.}~\bibnamefont {Koonin}}, \
  and\ \bibinfo {author} {\bibfnamefont {J.~W.}\ \bibnamefont {Negele}},\
  }\href {\doibase 10.1103/PhysRevC.13.1226} {\bibfield  {journal} {\bibinfo
  {journal} {Phys. Rev. C}\ }\textbf {\bibinfo {volume} {13}},\ \bibinfo
  {pages} {1226} (\bibinfo {year} {1976})}\BibitemShut {NoStop}%
\bibitem [{\citenamefont {Baran}\ \emph {et~al.}(2011)\citenamefont {Baran},
  \citenamefont {Sheikh}, \citenamefont {Dobaczewski}, \citenamefont
  {Nazarewicz},\ and\ \citenamefont {Staszczak}}]{Bar11}%
  \BibitemOpen
  \bibfield  {author} {\bibinfo {author} {\bibfnamefont {A.}~\bibnamefont
  {Baran}}, \bibinfo {author} {\bibfnamefont {J.~A.}\ \bibnamefont {Sheikh}},
  \bibinfo {author} {\bibfnamefont {J.}~\bibnamefont {Dobaczewski}}, \bibinfo
  {author} {\bibfnamefont {W.}~\bibnamefont {Nazarewicz}}, \ and\ \bibinfo
  {author} {\bibfnamefont {A.}~\bibnamefont {Staszczak}},\ }\href {\doibase
  10.1103/PhysRevC.84.054321} {\bibfield  {journal} {\bibinfo  {journal} {Phys.
  Rev. C}\ }\textbf {\bibinfo {volume} {84}},\ \bibinfo {pages} {054321}
  (\bibinfo {year} {2011})}\BibitemShut {NoStop}%
\bibitem [{\citenamefont {Thouless}\ and\ \citenamefont
  {Valatin}(1962)}]{TV62}%
  \BibitemOpen
  \bibfield  {author} {\bibinfo {author} {\bibfnamefont {D.~J.}\ \bibnamefont
  {Thouless}}\ and\ \bibinfo {author} {\bibfnamefont {J.~G.}\ \bibnamefont
  {Valatin}},\ }\href {\doibase 10.1016/0029-5582(62)90741-1} {\bibfield
  {journal} {\bibinfo  {journal} {Nucl. Phys.}\ }\textbf {\bibinfo {volume}
  {31}},\ \bibinfo {pages} {211} (\bibinfo {year} {1962})}\BibitemShut
  {NoStop}%
\bibitem [{\citenamefont {Bohr}\ and\ \citenamefont {Mottelson}(1975)}]{BM75}%
  \BibitemOpen
  \bibfield  {author} {\bibinfo {author} {\bibfnamefont {A.}~\bibnamefont
  {Bohr}}\ and\ \bibinfo {author} {\bibfnamefont {B.~R.}\ \bibnamefont
  {Mottelson}},\ }\href@noop {} {\emph {\bibinfo {title} {Nuclear Structure,
  Vol. II}}}\ (\bibinfo  {publisher} {W. A. Benjamin},\ \bibinfo {address} {New
  York},\ \bibinfo {year} {1975})\BibitemShut {NoStop}%
\bibitem [{\citenamefont {Hinohara}(2015)}]{Hino2015}%
  \BibitemOpen
  \bibfield  {author} {\bibinfo {author} {\bibfnamefont {N.}~\bibnamefont
  {Hinohara}},\ }\href {\doibase 10.1103/PhysRevC.92.034321} {\bibfield
  {journal} {\bibinfo  {journal} {Phys. Rev. C}\ }\textbf {\bibinfo {volume}
  {92}},\ \bibinfo {pages} {034321} (\bibinfo {year} {2015})}\BibitemShut
  {NoStop}%
\bibitem [{\citenamefont {Nakatsukasa}\ \emph
  {et~al.}(2016{\natexlab{b}})\citenamefont {Nakatsukasa}, \citenamefont
  {Matsuyanagi}, \citenamefont {Matsuzaki},\ and\ \citenamefont
  {Shimizu}}]{NMMS16}%
  \BibitemOpen
  \bibfield  {author} {\bibinfo {author} {\bibfnamefont {T.}~\bibnamefont
  {Nakatsukasa}}, \bibinfo {author} {\bibfnamefont {K.}~\bibnamefont
  {Matsuyanagi}}, \bibinfo {author} {\bibfnamefont {M.}~\bibnamefont
  {Matsuzaki}}, \ and\ \bibinfo {author} {\bibfnamefont {Y.~R.}\ \bibnamefont
  {Shimizu}},\ }\href {http://stacks.iop.org/1402-4896/91/i=7/a=073008}
  {\bibfield  {journal} {\bibinfo  {journal} {Physica Scripta}\ }\textbf
  {\bibinfo {volume} {91}},\ \bibinfo {pages} {073008} (\bibinfo {year}
  {2016}{\natexlab{b}})}\BibitemShut {NoStop}%
\end{thebibliography}%


\end{document}